\documentclass[pra]{revtex4}
\usepackage{amsmath,amssymb,mathrsfs}
\usepackage{psfrag}
\usepackage{graphicx}
\usepackage{graphics}
\usepackage{epsfig}
\usepackage{bm}
\usepackage{color}
\usepackage{verbatim,color,ulem}

\begin{document}

\title{Reliability of the Ginzburg-Landau Theory in the BCS-BEC Crossover 
by Including Gaussian Fluctuations for 3D Attractive~Fermions}

\author{F. Pascucci$^{1}$, A. Perali$^{1}$, and L. Salasnich$^{2}$}
\affiliation{$^{1}$School of Science and Technology, Physics Division, 
University of Camerino, Via Madonna delle Carceri 9B, 62032 Camerino, Italy\\
$^{2}$Dipartimento di Fisica e Astronomia ``Galileo Galilei'', 
University of Padova, Via Marzolo 8, Padova, Italy}

\begin{abstract}
We calculate the parameters of the Ginzburg--Landau (GL) equation of a three-dimensional attractive Fermi gas around the superfluid critical temperature. We compare different levels of approximation throughout the Bardeen--Cooper--Schrieffer (BCS) to the Bose--Einstein Condensate (BEC) regime. We show that the inclusion of Gaussian fluctuations strongly modifies the values of the Ginzburg--Landau parameters approaching the BEC regime of the crossover. We investigate the reliability of the Ginzburg--Landau theory, with fluctuations, studying the behavior of the coherence length and of the critical rotational frequencies throughout the BCS-BEC crossover. The effect of the Gaussian fluctuations gives qualitative correct trends of the considered physical quantities from the BCS regime up to the unitary limit of the BCS-BEC crossover. Approaching the BEC regime, the Ginzburg--Landau equation with the inclusion of Gaussian fluctuations turns out to be unreliable.
\end{abstract}

\maketitle

\section{Introduction}
The last two decades of developments in the confinement, cooling, and control of the interaction in alkali-metal atomic gases have powered the interest in the BCS-BEC crossover~\cite{Hu2006, Diener2008}. The crossover from the BCS state to the BEC one has been observed in two-hyperfine-component Fermi gases of $^{40}$K atoms and $^6$Li atoms~\cite{Regal2004, Kinast2004} with the use of Fano--Feshbach resonances~\cite{Inouye1998}. A simple, but powerful tool to study superconductivity and superfluidity around the critical temperature is the Ginzburg--Landau theory~\cite{Ginzburg1950}. In \cite{Banerjee2011}, \mbox{Banerjee et al.} showed that at the phenomenological level, the thermodynamic properties of cuprate superconductors can be described through a free-energy functional having the GL form for a wide range of temperatures and hole dopings. An extension of the GL theory to multicomponent systems was exploited in \cite{Salasnich2019, Saraiva2020} to characterize the screening of detrimental fluctuations in superconductors having coexisting shallow and deep electronic bands. In addition, in \cite{Milosevic2020}, Milo$\hat s$evi$\acute c$ et al. showed how the GL theory is suitable to describe different systems as hybrid and layered superconductors, which exhibit anisotropy in different directions. The main motivation of our work rests on the formulation of an alternative, simpler approach for the calculation of complex superfluid properties throughout the BCS-BEC crossover close to the critical temperature. Indeed, the microscopic study of collective modes~\cite{Combescot2006}, Josephson effect~\cite{Pascucci2020}, vortex configurations, and multicomponent superfluidity~\cite{Salasnich2019, Saraiva2020} requires in general demanding numerical calculations. Here, we propose that, through the GL theory with fluctuations, it is possible to study several superfluid quantities by simply knowing the value of the GL parameters. To investigate the reliability of the GL theory with fluctuations throughout the BCS-BEC crossover, we compared the result obtained from the GL theory with the microscopic approach for the coherence length and the critical rotational frequencies.

In this work, we studied a three-dimensional neutral fermionic system close to the superfluid phase transition temperature. We derived the phenomenological Ginzburg--Landau equation through the Gorkov's microscopic approach~\cite{Gorkov1959}. In particular, we calculated the Ginzburg--Landau parameters in the BCS-BEC crossover 
{through the integral functional approach as in~\cite{DeMelo1993, Drechsler1992, Micnas2001}}{extended the procedure reported in Ref.~\cite{DeMelo1993}}. The results obtained applying the mean-field approximation were compared with the ones applying the beyond-mean-field approximation. Through GL parameters, we investigated the behavior of the coherence length, comparing our results with the one obtained by Palestini and Strinati through the microscopic diagrammatic approach in~\cite{Palestini2014}.
Then, we analyzed the critical rotational frequencies that are the neutral system analogue of the critical magnetic fields for superconductors~\cite{Falk66}. 

\section{Methods}
The Hamiltonian density of the fermionic system considered in our work is given by:
\begin{equation}
H(x)=\overline{\Psi}_{\sigma}(x)\Bigg[-\frac{\hbar^2\nabla^2}{2m}-\mu\Bigg]\Psi_{\sigma}(x)-g\overline{\Psi}_{\uparrow}(x)\overline{\Psi}_{\downarrow}(x)\Psi_{\uparrow}(x)\Psi_{\downarrow}(x),
\label{densham}
\end{equation}
where $m$ is the mass of the fermions, $\mu$ is the chemical potential, and $\Psi_{\sigma}$ and $\overline{\Psi}_{\sigma}$ are the complex Grassmann fields with spin $\sigma=\uparrow, \downarrow$. This Hamiltonian density describes a system with a single-channel interaction, where $g>0$ is the strength of the s-wave interatomic coupling. Through a functional integral formulation, it is possible to study the finite-temperature BCS-BEC crossover tuning the coupling $g$. Introducing the bosonic field $\Delta(x)$:
\begin{equation}
\Delta(x)=g<\Psi_{\downarrow}(x)\Psi_{\uparrow}(x)>,
\end{equation}
and applying the Hubbard--Stratonovich transformation the Hamiltonian density reads:
\vspace{3pt}

\nointerlineskip
\begin{equation}
H(x)=\overline{\Psi}_{\sigma}(x)\Bigg[-\frac{\hbar^2\nabla^2}{2m}-\mu\Bigg]\Psi_{\sigma}(x)+\\
\frac{|\Delta(x)|^2}{g}-\overline{\Delta}(x)\Psi_{\downarrow}(x)\Psi_{\uparrow}(x)-\Delta(x)\overline{\Psi}_{\uparrow}(x)\overline{\Psi}_{\downarrow}(x),
\label{H}
\end{equation}
where $\Delta(x)$ is the gap energy. Following~\cite{DeMelo1993}, we proceed with the integral formulation of the effective action $S$:
\begin{equation}
S_{eff}[\Delta(x)]=\int_{0}^{\beta}d\tau \int_{}^{}dx \frac{|\Delta(x)|^2}{g}-Tr\Bigl[ ln G^{-1}[\Delta(x)]\Bigr],
\end{equation}
where $\beta=1/k_BT$ ($k_B$ is the Boltzmann constant), $T$ the temperature, and $G^{-1}[\Delta(x)]$ the inverse of the Nambu propagator:
\begin{equation}
G^{-1}(x,x')=\begin{bmatrix}-\partial_{\tau}+\frac{\nabla^2}{2m}+\mu&\Delta(x)\\\overline{\Delta}(x)&-\partial_{\tau}-\frac{\nabla^2}{2m}-\mu\end{bmatrix}\delta(x-x').
\label{G}
\end{equation}

The trace in $S_{eff}$ is over space $\vec{x}$, imaginary time $\tau$, and Nambu indices. We can move to the momentum space and expand the effective action in terms of powers of $\Delta(q)$ since the gap energy is small around the critical temperature. The effective action can be written~as:
\begin{equation}
S_{eff}[\Delta,\overline{\Delta}]=\frac{1}{V}\sum_q\frac{|\Delta(q)|^2}{\Pi(q)}+\frac{1}{2}\frac{1}{V}\sum_{q_1,q_2,q_3}b\Delta_{q_1}\Delta_{q_2}^*\Delta_{q_3}\Delta_{q_1-q_2+q_3}^*+.....
\label{Seff}
\end{equation}
where $\Pi$ is the coefficient of the second-order terms $|\Delta(q)|^2$ with all the gradient orders of $\Delta(q)$ and $V$ is the volume of the system. To obtain a relation formally equal to the GL functional, we have to consider only the quadratic term, the quadratic gradient term, and the quartic term of Equation~\eqref{Seff}. In this approximation, the inverse of the coefficient $\Pi$~reads:
\begin{equation}
\Pi^{-1}(q)=a+\frac{c|q|^2}{2m}.
\label{Pi}
\end{equation}

The parameters of the expansion terms have the following form~\cite{DeMelo1993}:
\begin{gather}
a=-\frac{m}{4\pi\hbar^2 a_F}+\frac{1}{V}\sum_k\Bigg[\frac{1}{2\epsilon_k}-\frac{tanh(\beta\xi_k/2)}{2\xi_k}\Bigg],
\label{adelta}
\\
b=\frac{1}{V}\sum_k\Bigg[\frac{tanh(\beta\xi_k/2)}{4\xi^3_k}-\frac{\beta sech^4(\beta\xi_k/2)}{8\xi_k^2}\Bigg],
\label{bdelta}
\\
c=\frac{1}{V}\sum_k\Bigg[\frac{tanh(\beta\xi_k/2)}{4\xi^2_k}-\frac{\beta sech^2(\beta\xi_k/2)}{8\xi_k}\Bigg],
\label{cdelta}
\end{gather}
where $\xi_k=\epsilon_k-\mu$ with $\epsilon_k=\hbar^2k^2/2m$. Minimizing the effective action S$_{eff}$ in the real space, we obtain the GL equation:
\begin{gather}
\frac{\delta S_{eff}[\Delta,\overline{\Delta}]}{\delta\overline{\Delta}}=0,\\
\Bigg[a+b|\Delta(x,t)|^2-\frac{\hbar^2c}{2m}\nabla^2\Bigg]\Delta(x,t)=0.
\end{gather}

The GL equation is generally formulated with the Cooper pair field $\Psi(x,t)$ instead of the gap energy $\Delta(x,t)$. It is possible to shift from one notation to the other replacing $\Delta(x,t)=\Psi(x,t)/\sqrt{2c}$. In this way, we obtain:
\begin{equation}
\Bigg[A+\frac{B}{2}|\Psi(x,t)|^2-\Gamma\nabla^2\Bigg]\Psi(x,t)=0,
\label{gle}
\end{equation}
where:
\begin{gather}
A=\frac{a}{2c},
\label{apsi}
\\
B=\frac{b}{2c^2},
\label{bpsi}
\\
\Gamma=\frac{\hbar^2}{4m}.
\label{cpsi}
\end{gather}

\subsection{Gap Equation and Number Equation}
The GL parameters depend on the temperature $T$, the chemical potential $\mu$, and the interaction strength $g$. To determine this triad of values throughout the BCS-BEC crossover, we start solving the gap and the number equation at the lowest-mean-field level of approximation at the critical temperature $T_{c0}$. The first one is calculated by applying the saddle point condition:
\begin{equation}
\frac{\partial S_{eff}(\Delta(x))}{\partial\Delta(x)}=0.
\end{equation}

In this way, we obtain the linearized BCS equation for the critical temperature:
\begin{equation}
\frac{1}{g}=\frac{1}{V}\sum_{k} \frac{tanh(\xi_k/2T_{c0})}{2\xi_k}.
\end{equation}

Replacing the summation over the discretized wave vectors $k$ with the integral over continuous wave vectors, an ultraviolet divergence appears in the equation for $T_{c0}$. This divergence can be eliminated by introducing a regularization based on the scattering length $a_F$, defined by the two-body problem via $m/4\hbar^2\pi a_F=-1/g+\sum_k(\epsilon_k)^{-1}$. The regularized equation for $T_{c0}$ reads:
\begin{equation}
-\frac{m}{4\pi\hbar^2 a_F}=\frac{1}{V}\sum_{k}\Bigg[\frac{tanh(\xi_k/2T_{c0})}{2\xi_k}-\frac{1}{2\epsilon_k}\Bigg].
\label{gap}
\end{equation}

The number equation is given by:
\begin{equation}
n=-\frac{1}{V}\frac{\partial\Omega[\Delta(x)]}{\partial\mu},
\end{equation}
where $n$ is the total fermionic density of the system, which in our case is fixed, and $\Omega$=$-ln(Z)/\beta$ is the grand potential where $Z=\int D\Delta D\overline{\Delta}exp(-S_{eff}[\Delta,\overline{\Delta}])$ is the partition function. At the mean-field approximation, we impose $\Delta(x)=0$. In this way, one obtains $Z=e^{-S_0}$ and $\Omega_0=S_0/\beta$. The number equation is:
\begin{equation}
n=n_0(\mu_{c0},T_{c0})=\frac{1}{V}\sum_{k}\Bigg[1-tanh\Bigg(\frac{\xi_k}{2T_{c0}}\Bigg)\Bigg]=2 f(E_k),
\label{number}
\end{equation}
where $f(E_k)$ is the Fermi distribution function. Comparing Equation (19) with Equation~(8), we can see that the GL parameter $a$ is zero at the critical temperature. What is generally done is an expansion of the parameter around the critical temperature:
\begin{equation}
a=\frac{1}{V}\frac{da}{dT}\Bigg|_{T_{c0}}(T-T_{c0})=\alpha (T-T_{c0}),
\end{equation}
where $\alpha$ is the derivative of $a$ with respect to $T$ evaluated at the critical temperature, and it~reads:
\begin{equation}
\alpha=\frac{1}{V}\sum_k\frac{sech^2(\xi_k/2k_BT_{c0})}{4k_BT_{c0}^2}.
\end{equation}

The same happens with the GL parameter $A$ defined by Equation (13):
\begin{equation}
A=\frac{dA}{dT}\Bigg|_{T_{c0}}(T-T_{c0})=\overline{A}(T-T_{c0}),
\end{equation}
where $\overline{A}$ is the derivative of the GL parameter $A$ evaluated at the critical temperature. Solving Equations~\eqref{gap} and \eqref{number}, we can estimate the saddle point T$_{c0}$ and $\mu_{c0}$, as a function of the scattering length $a_F$. We get a T$_{c0}$ that grows continuously from BCS to BEC~\cite{DeMelo1993}. It is indeed well known that the mean-field approximation is not enough to properly describe the system in the BEC regime at finite temperature. We can further explore this problem using the Ginzburg--Levanyuk criterion~\cite{Larkin2005a}. The mean-field approximation is valid only if the temperature T of the system satisfies the inequality:
\begin{equation}
\frac{T-T_{c0}}{T_{c0}}>Gi_{3D},
\label{Giwindow}
\end{equation}
where $Gi_{3D}$ is the Ginzburg--Levanyuk number in 3D:
\begin{equation}
Gi_{3D}=\Bigg(\frac{BT_{c0}^{1/2}}{8\pi\Gamma^{3/2}\overline{A}^{3/2}}\Bigg)^2.
\label{Gieq}
\end{equation}

{Equations (25) and (26) provide the temperature interval around the critical one in which fluctuations cannot be neglected throughout the BCS-BEC crossover.
Replacing \mbox{Equations (8)--(10)}} in Equations (14)--(16) and then in Equation \eqref{Gieq}, we study the behavior of $Gi_{3D}$ along with the BCS-BEC crossover.

\begin{figure}[t]
\includegraphics[width=9cm]{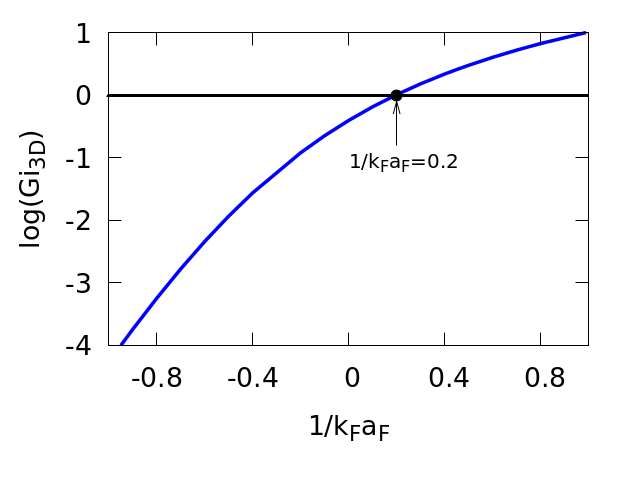}
\caption{{The Ginzburg--Levanyuk} number $Gi_{3D}$ as a function of the inverse normalized scattering length $1/k_Fa_F$ (blue line) and the reference value $Gi_{3D}=1$ (black line). Here, $k_F=3\pi^2n^{1/3}$ is the noninteracting Fermi wave vector.}
\label{Gi}
\end{figure}

As we can see from Figure~\ref{Gi}, around the unitary limit, fluctuations cannot be neglected, while in the BCS regime, the mean-field approximation works very well near the critical temperature. For these reasons, in the BCS-BEC crossover regime, it is necessary to include beyond-mean-field Gaussian fluctuations around the saddle point. The Gaussian action expanded to second order in $\Delta(x)$ is given by:
\begin{equation}
S_{Gauss}=S_{0}+\beta V\sum_{q,\omega_l}\Pi^{-1}(q,i\omega_l)|\Delta(q,i\omega_l)|^2,
\label{Sgauss}
\end{equation}
where $\Pi^{-1}$ is the same as Equation \eqref{Seff} and $\omega_l=2l\pi/\beta$ is the bosonic Matsubara frequency. We include all the second-order terms of the gradient expansion. The partition function $Z=\int D[\Delta,\overline{\Delta}]e^{-S_{Gauss}}$ reads:
\begin{equation}
Z_{Gauss}=e^{-S_0}det\Big[\Pi^{-1}\Big]^{-1}
\end{equation}
and the grand potential $\Omega_{Gauss}=-ln[Z_{Gauss}]/\beta$ reads:
\begin{equation}
\Omega_{Gauss}=\Omega_0+\frac{ln\Big[det(\Pi^{-1})\Big]}{\beta}.
\end{equation}

For a nonsingular square matrix $A$, we have $log\Big[det(A)\Big]=Tr\Big[log(A)\Big]$. In this way, we obtain the beyond-mean-field grand potential:
\begin{equation}
\Omega_{Gauss}=\Omega_0+\frac{Tr\Big[ln\Big(\Pi^{-1}\Big)\Big]}{\beta}
\end{equation}
and the beyond-mean-field number equation:
\begin{equation}
n=-\frac{1}{V}\frac{\partial\Omega_{Gauss}}{\partial \mu},
\end{equation}
which yields:
\begin{equation}
n=n_0+\frac{k_BT_c}{V}\frac{\partial Tr[ln(\Pi)]}{\partial\mu},
\end{equation}
where $n_0$ is given by Equation~\eqref{number}. Following the Nozières--Schmitt-Rink approach~\cite{NSR}, we can rewrite Equation~(32) in terms of the phase shift defined by $\Pi(q,\omega_q\pm i0^+)=|\Pi(q,\omega)|$ $exp[\pm i\delta(q,\omega)]$. The number equation incorporating the effects of Gaussian fluctuations is:
\begin{equation}
n=n_0(\mu_c,T_c)+\sum_q\int_{-\infty}^{\infty}\frac{d\omega}{\pi}n_B(\omega)\frac{\partial\delta(q,\omega)}{\partial\mu},
\label{nbmf}
\end{equation}
where $n_B(\omega)=1/[exp(\beta\omega)-1]$ is the Bose--Einstein distribution function and $\delta(q,\omega)=Arctan[Im[\Pi(q,\omega)^{-1}]/Re[\Pi(q,\omega)^{-1}]]$ is the phase shift. As discussed and first obtained in~\cite{DeMelo1993}, we solve Equations~\eqref{gap} and \eqref{nbmf} with $T_c$ instead of $T_{c0}$ in Equation~\eqref{gap} to obtain the critical temperature $T_c$ and the critical chemical potential $\mu_c$ in the BCS-BEC crossover in the beyond-mean-field approximation.

\begin{figure}[t]
\includegraphics[width=12 cm]{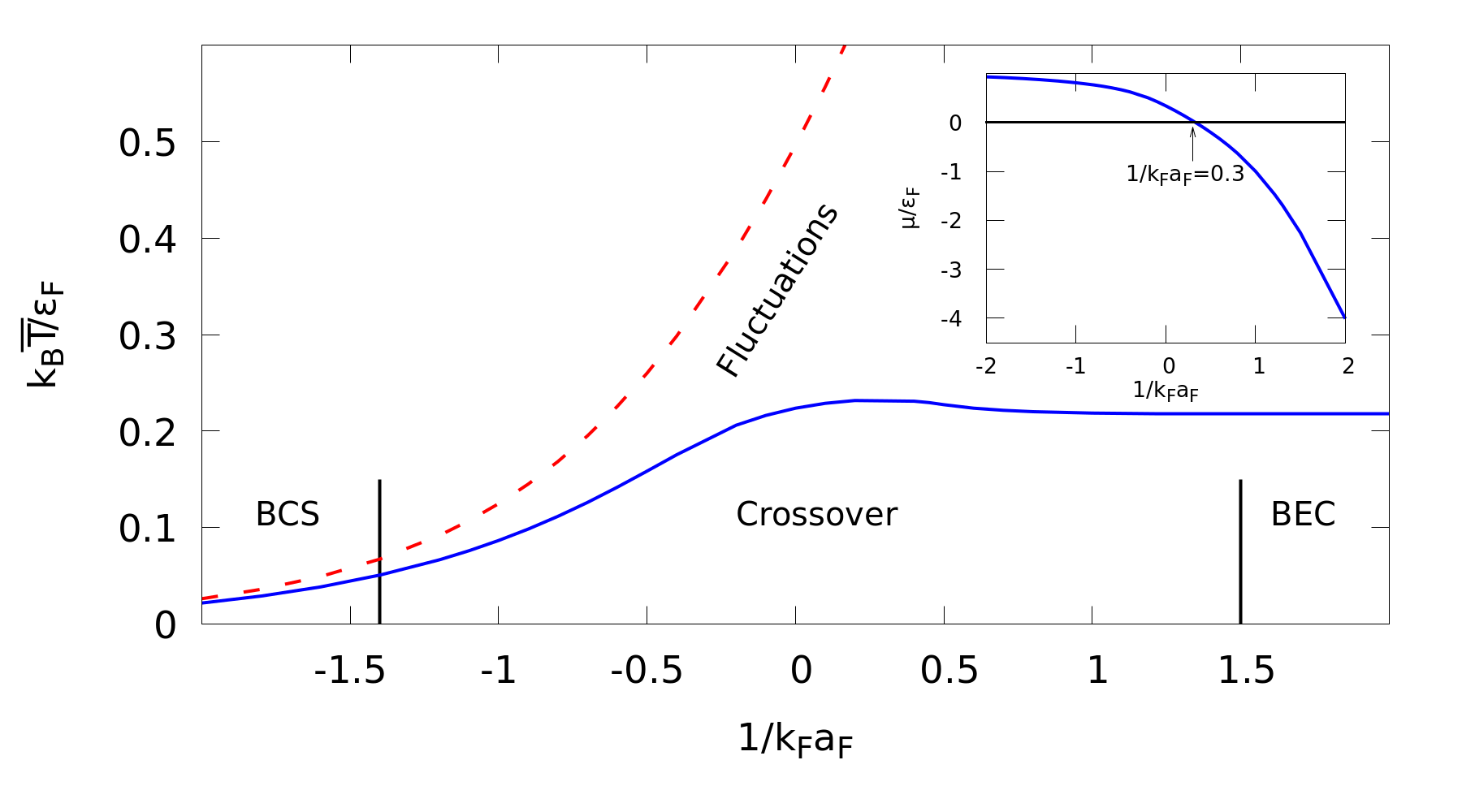}
\caption{{The normalized critical} temperature $k_B\overline{T}/\epsilon_F$ as a function of $1/k_Fa_F$. The red dashed line ($\overline{T}=T_{c0}$) is the mean-field result, while the blue solid line ($\overline{T}=T_{c}$) is the beyond-mean-field result. In the inset is reported the normalized critical chemical potential $\mu_c/\epsilon_F$ in the beyond-mean-field approximation as a function of $1/k_Fa_F$. Here, $\epsilon_F$ is the Fermi energy of the noninteracting system.}
\label{tc}
\end{figure}

In Figure~\ref{tc}, we report the behavior of the superfluid critical temperature in the two approximations. The two black vertical lines delimitate the region in which the system is considered in the BCS, crossover, or BEC regime.
{The behavior of the normalized chemical potential is reported in the inset of Figure~\ref{tc}. At the unitary limit, $1/k_Fa_F=0$, $\mu/\epsilon_F$ is positive, meaning that the system is still in the crossover regime of the BCS-BEC crossover with the Fermi surface that survives. To enter the BEC regime, it is necessary to increase the coupling beyond $1/k_Fa_F=0.3$, where the chemical potential becomes negative and the Fermi surface collapses.}
In~\cite{Strinati93}, Pistolesi and Strinati, studying the behavior of the chemical potential as a function of the product $k_F \xi$, where $\xi$ is the intrapair coherence length obtained from the pair correlation function, found that the system can be considered in the crossover region when $\pi^{-1}<k_F \xi<2\pi$. In~\cite{Palestini2014}, Palestini and Strinati investigated the behavior of the coherence length throughout the BCS-BEC crossover using a diagrammatic approach. In this way, we find that the system can be considered in the crossover regime when $-1.4<1/k_Fa_F<1.5$. As shown in Figure~\ref{tc}, the effect of the fluctuations can be neglected in the BCS regime, but they become relevant already at the unitary limit and are fundamental to obtain the BE condensation critical temperature $k_BT_c\simeq0.218\epsilon_F$ for noninteracting bosons in the strong-coupling regime.

\subsection{Ginzburg--Landau Parameters and Characteristic Quantities}
\begin{figure}[t]
\includegraphics[width=11.8 cm]{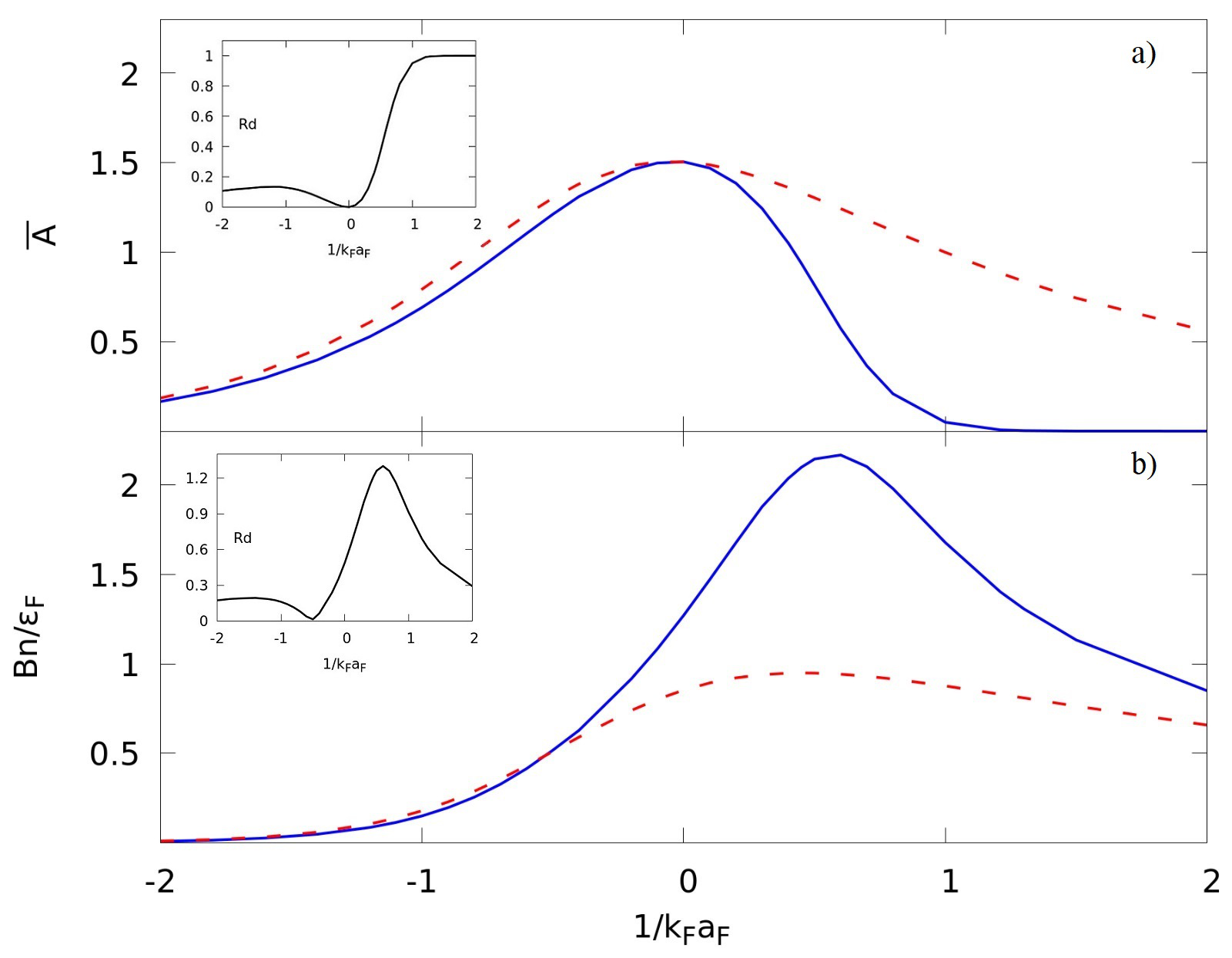}
\caption{{The GL parameters} $\overline{A}$ (\textbf{a}) and $B$ (\textbf{b}) as a function of $1/k_Fa_F$ in the mean-field (red dashed line) and beyond-mean-field (blue solid lined) approximation. In the insets is reported the relative difference $R_d$ between the mean-field and the beyond-mean-field results.}
\label{GLp}
\end{figure}

In the formulation of the GL functional, we found that the coefficient of the quartic term $\Gamma$ is constant throughout the BCS-BEC crossover. We analyzed the behavior of the coefficients $\overline{A}$ and $B$ in the mean-field and beyond-mean-field approximation at their respective critical temperatures $T_{c0}$ and $T_c$. As we can see from Figure~\ref{GLp}a for the GL parameter $\overline{A}$, Gaussian fluctuations are negligible in the BCS regime and relevant approaching the BEC regime, but as reported in the inset, the relative difference $R_d$ between the two approximations does not grow continuously: there is a minimum around the unitary limit. Instead, for the GL parameter $B$, reported in Figure~\ref{GLp}b, fluctuations are also relevant at the unitary limit with a maximum for $R_d$ around the value $1/k_Fa_F\simeq0.5$. From the inset in Figure~\ref{GLp}b, we can also see a minimum for the relative difference at $1/k_Fa_F=-0.5$.

\vskip 0.5cm 

To better understand these results, with the GL theory, 
we study a fundamental length scale of the superfluid state: 
the coherence length, which as a function of the GL parameters, 
can be written as~\cite{Tinkham2004}:
\begin{equation}
\xi_{GL}=\sqrt{-\frac{\Gamma}{A(T)}}.
\label{eqxi}
\end{equation}

\begin{figure}[t]
\includegraphics[width=10 cm]{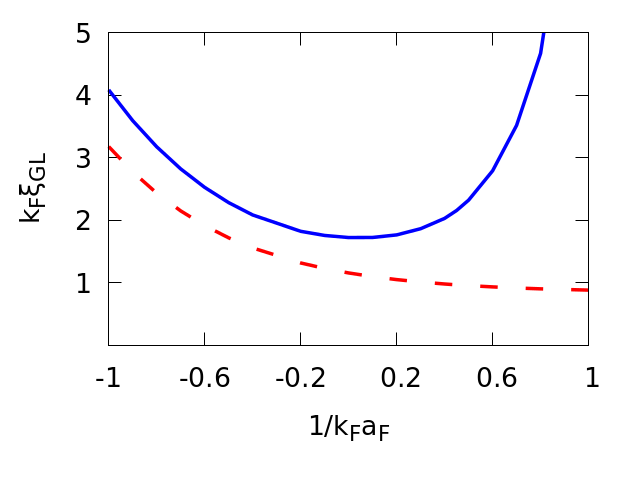}
\caption{{The normalized Ginzburg--Landau} coherence length $k_F\xi_{GL}$ as a function of $1/k_Fa_F$ in the mean-field approximation (red dashed line) at temperature $T=0.5T_{c0}$ and in the beyond-mean-field approximation (blue solid line) at temperature $T=0.5T_c$.}
\label{xitc}
\end{figure}

The coherence length is the characteristic length of the spatial variation of the Cooper pair wave function, quantifying interpair correlations. In their work~\cite{Palestini2014}, Palestini and Strinati, using a diagrammatic approach, found that in the BEC regime at zero temperature, the difference between the mean-field and the beyond-mean-field results is very small. This result can be reasonably extended also at finite temperature since they showed that until $T=0.5T_c$, the coherence length weakly depends on the temperature. In~\cite{Pieri2004}, \mbox{Pieri {et al.}} showed in Figure~\ref{xitc} that the chemical potential $\mu$ is basically constant for $0<T<T_c$. For these reasons, we can reasonably extend the use of the Ginzburg--Landau theory down to $0.5T_c$. In Figure~\ref{xitc}, we can see that at $T=0.5T_c$, the difference between the mean-field and the beyond-mean-field $\xi_{GL}$ increases as we move into the BEC regime. Instead, in the BCS and crossover regime (BCS side), up to $1/k_Fa_F=0.5$, we obtain a coherence length increased by fluctuations with respect to the mean-field of the right order of magnitude, in agreement with the findings in \cite{Palestini2014}.

In the BEC regime, the increasing discrepancy is linked to how we treated fluctuations of the order parameter. To formulate the GL functional, we made a series expansion of the effective action in the small-order parameter, and then we took the zero-order and the second-order term of $\Delta^2(q)$. Crossing over from the BCS to the BEC regime, since the binding energy of the Cooper pairs increases, it is reasonable to think that, in the strongly interacting limit, the fluctuations become more and more important, and it is no more sufficient to stop the second-order expansion to describe their effect properly. Indeed, as we showed in the previous section, it is necessary to consider all the orders of the gradient term to obtain the right critical temperature in the BEC regime. Generally, it is important to study the healing length, because in superconductors, it is related to the spatial variation of the order parameter inside an Abrikosov vortex close to the critical temperature~\cite{Tinkham2004}. In our case of a neutral fermionic system, we can study superfluid characteristic quantities: the critical rotational frequencies. For charged superconductors in a magnetic field, there exists a critical value $H_{c1}$ over which superconducting vortices start to appear in the system. The analog for a neutral system is the critical rotational frequency. For a rotating superfluid with angular frequency $\omega$, there exists a critical value $\omega_{c1}$ over which vortices form in the superfluid. Increasing $\omega$, as for the superconductors, more and more vortices populate the system until we reach a second critical rotational frequency $\omega_{c2}$, at which the vortices come into contact and the superfluidity is destroyed. This is the superconducting analog of the upper critical magnetic field $H_{c2}$. An estimation of $\omega_{c1}$ at which the first vortex enters the superfluid~\cite{Landau1981} is given by the relation:
\begin{equation}
\omega_{c1}=\frac{\hbar}{m}\frac{1}{R^2} ln\Bigg(\frac{R}{\xi}\Bigg),
\label{omegac1}
\end{equation}
where $R$ is the radius of the superfluid and $\xi$ is the healing length of the vortex. Equation~\eqref{omegac1} multiplied by $\hbar$ represents the energy of an isolated vortex, and it is identical to the one for $H_{c1}$ in which $\hbar/m$ is replaced by the magnetic quantum flux $\phi_0/2\pi$ and $R$ by the penetration length of the magnetic field $\lambda$. It is interesting to note that the length scales considered in the superconducting and superfluid case are very different. The penetration length $\lambda$ is orders of magnitude smaller than the superfluid radius $R$. This could be explained by considering the different nature of the vortices between the superconductive and superfluid case. In the first one, we have an external potential that tries to penetrate a superconductor, and over a certain value $H_{c1}$, it succeeds by the formation of superconducting vortices. Increasing the magnetic field, $H_{c2}$ is reached and superconductivity is destroyed: there are no more vortices, and the system is in the normal state. For a rotating superfluid, over $\omega_{c2}$, the superfluidity disappears, but the entire fluid is rotating as a unique vortex of radius $R$. The length in the numerator of the logarithm in Equation~\eqref{omegac1} could be related to the maximum size reachable for the vortex. Replacing Equation~\eqref{eqxi} in Equation~\eqref{omegac1}, we obtain a GL parameter-dependent relation for the critical rotational frequency $\omega_{c1}$. To study experimentally a Fermi gas, used an external magnetic field is generally to confine the system in a specific region. The effect of the trapping potential on a rotating superfluid throughout the BCS-BEC crossover was studied in~\cite{Song2012, Machida2006}. They found that the presence of a trapping potential induces a Landau-level structure in the energy spectrum when the superfluid rotational frequency approaches the trapping potential rotational frequency $\omega_r$. In particular, in the BEC regime, since the chemical potential is less than the cyclotron energy gap ($2\hbar\omega_r$), the system enters the lowest Landau level, while at unitary and in the BCS regime, where the chemical potential is large compared with $2\hbar\omega_r$, it occupies many low-lying Landau levels. In this work, we did not consider the presence of the trapping potential because, as we saw, with the GL theory, we cannot properly describe the superfluid in the BEC regime, while in the BCS regime, the discretization of the energy spectrum in Landau levels can be neglected.

Continuing to increase the frequency of rotation above $\omega_{c1}$, more and more vortices appear in the superfluid until they came into contact at $\omega_{c2}$. For a superconducting system, the upper critical magnetic filed $H_{c2}$ is given by~\cite{Tinkham2004}:
\begin{equation}
H_{c2} = \frac{\phi_0}{2\pi}\frac{1}{\xi^2}.
\end{equation}

Replacing $\phi_0/2\pi$ with $\hbar/m$ as done above, we obtain the relation for the upper critical rotational frequency $\omega_{c2}$:
\begin{equation}
\omega_{c2}=\frac{\hbar}{m}\frac{1}{\xi^2}.
\label{eqomegac2}
\end{equation}

\begin{figure}[t]
\includegraphics[width=10.0 cm]{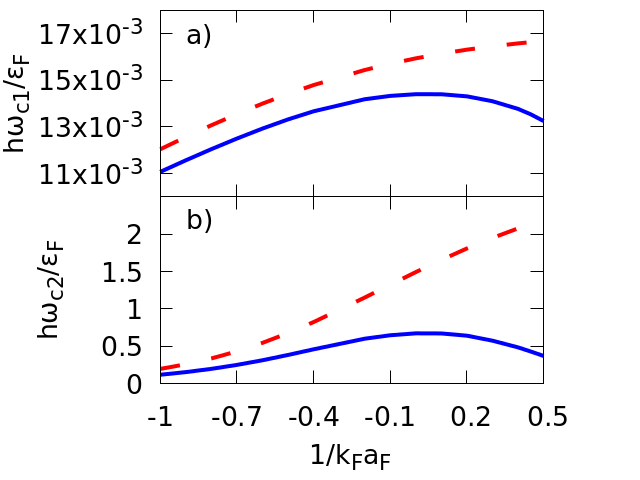}
\caption{{The normalized critical }rotational frequency $\hbar\omega_{c1}/\epsilon_F$ (\textbf{a}) and $\hbar\omega_{c2}/\epsilon_F$ (\textbf{b}) as a function of $1/k_Fa_F$ in the mean-field approximation (red dashed line) at the temperature $T=0.5T_{c0}$ and in the beyond-mean-field approximation (blue solid line) at the temperature $T=0.5T_c$.}
\label{omegatc}
\end{figure}

In Figure~\ref{omegatc}a, we report the behavior of the lower critical rotational frequency $\omega_{c1}$ throughout the BCS-BEC crossover setting $Rk_F=72$ as in~\cite{Simonucci2015}, and in Figure~\ref{omegatc}b, we report the behavior of the upper critical rotational frequency. The coherence length in Equations (35) and (37) is taken to be the Ginzburg--Landau coherence length, as studied and evaluated in this work. As we can see, $\omega_{c2}$ is two orders of magnitude larger than $\omega_{c1}$ in both approximations. The GL coherence length can also be related to the superfluid critical velocity $v_c$~\cite{Combescot2006}:
\begin{equation}
\xi_{GL}=\frac{\hbar}{mv_c}.
\label{vc}
\end{equation}

Replacing Equation~\eqref{eqxi} in Equation~\eqref{vc}, we obtain a relation for the critical velocity as a function of the GL parameters, while usually, the critical velocity is obtained using the Landau criterion, as in~\cite{Combescot2006}. Instead of the cusp in the crossover point in which the pair-breaking and the bosonic branch of the Landau criterion intersect each other, we have a maximum in the critical velocity that continuously describes the crossover region in which there is a competition between the fermionic and the bosonic behavior of the fluctuating Cooper pairs. In~\cite{Combescot2006}, the critical velocity in the BEC regime went to zero much slower than in the BCS regime. The beyond-mean-field results did not follow this behavior because, as we showed for the coherence length, in the BEC regime, the Ginzburg--Landau theory is not reliable.

\section{Conclusions}
The Ginzburg--Landau equation was obtained through a series expansion of the effective action in powers of the order parameter, keeping the quadratic, gradient, and quartic terms. We studied the behavior of the Ginzburg--Landau parameters around the superfluid critical temperature throughout the BCS-BEC crossover by solving the gap equation and the number equation in the mean-field and beyond-mean-field approximation. The number equation was obtained starting from an effective action formed by the mean-field and the Gaussian fluctuation terms, considering all second-order terms of the expansion. We found that the effects of the Gaussian fluctuations on the Ginzburg--Landau parameters $\overline{A}$ and $B$ are relevant, in particular in the unitary and BEC regime of the BCS-BEC crossover. We tested the reliability of the Ginzburg--Landau theory by studying the coherence length, written as a function of the Ginzburg--Landau parameters, and compared our results to the ones obtained by the diagrammatic approach in~\cite{Palestini2014}. We found a reasonable agreement in the range of couplings from the BCS regime up to $1/k_Fa_F=0.5$. Beyond this value of $(k_Fa_F)^{-1}$, entering the BEC regime, the effect of fluctuations becomes too strong and the Ginzburg--Landau equation is not reliable. Increasing the strength of the interatomic attraction, the Cooper pair size decreases, and the characteristic length of the pair wave function becomes comparable to the characteristic length of the pair fluctuations. To obtain the correct results, it is necessary to include gradients of order higher than the second in the effective energy expansion. This is supported by the fact that at zero temperature, as shown in \cite{Huang2009}, the Ginzburg--Landau equation is well defined and becomes the Gross--Pitaevskii equation with a scattering length $a_B=2a_F$, while from the few-body calculations, we know that $a_B=0.6a_F$~\cite{Petrov2004}. In~\cite{Huang2009}, from the beyond-mean-field number equation of the Bogoliubov theory, in which the coefficient of the second-order term is not truncated as in Equation~\eqref{Pi}, they derived a new Gross--Pitaevskii equation with $a_b=0.56a_F$, which is very close to the correct result. Finally, we studied the upper and lower critical rotational frequency, written as a function of the Ginzburg--Landau parameters. We also compared our results to the one of the critical velocity in~\cite{Combescot2006}. We showed how the Gaussian fluctuations from the BCS to the unitary limit lead to a qualitative correct effect on the critical rotational frequencies and critical velocity. On the other hand, also in this case, we found that the Ginzburg--Landau equation is not able to describe the properties of the system in the BEC regime of the BCS-BEC crossover.
{In summary, we showed that the Ginzburg--Landau theory is a reliable tool to describe 3D systems formed by fermions interacting by a contact potential in the weak and intermediate coupling regime. Further investigations are needed to extend the results of our work to other interesting systems and phenomena, for example a Fermi system in the presence of impurity scattering processes~\cite{Palestini2013} or in the analysis of the Josephson effect~\cite{Spuntarelli2007} throughout the BCS-BEC crossover.}

\end{document}